# Search for chargino and neutralino associated production at the Tevatron

**Daniela Bortoletto**[*]

*Purdue University*
*525 Northwestern Avenue, West Lafayette I 47907-203, USA*
*E-mail:* `daniela@physics.purdue.edu`
Representing the CDF and D0 collaborations

I present a review of the searches for charginos and neutralinos using data from proton-antiproton collision at the centre of mass energy of 1.96 TeV collected by the CDF and the D0 experiments at the Tevatron during run II.

---

[*] Speaker

## 1. Introduction

The Standard Model of particles and interactions (SM) has been extraordinarily successful in describing experimental results. Nonetheless the SM is considered a low energy effective theory since it does not explain questions such as the unification of the forces of nature, the stability of the lightest Higgs boson mass under radiative corrections, and the nature of dark matter in the universe. One of the most promising models for physics beyond the Standard Model is Supersymmetry (SUSY) [1]. SUSY introduces a new particle for each SM particle with the same quantum numbers but the spin. Therefore if S is the spin, B is the Baryon number and L is the lepton number, SM and SUSY particles carry $R_{parity}= (-1)^{S+3B+L}$ equal to +1 and -1 respectively. Since SUSY particles remain to be observed, the SUSY symmetry must be broken. The minimal gravity mediated supersymmetry breaking (mSUGRA) with R-parity conservation is one of the most popular SUSY models. In this scenario $\tilde{\chi}_1^0$, the lightest neutralino, is the Lightest Supersymmetric Particle (LSP), an excellent candidate for the dark matter in the universe [2]. The SUSY parameter space in mSUGRA is usually studied in terms of the common scalar mass $m_0$, the common gaugino mass $m_{1/2}$, the ratio of the Higgs expectation values tan$\beta$, the sign of the Higgsino mass parameter $\mu$, and the common trilinear scalar coupling $A_0$.

Here I report about the latest results on the search for $\tilde{\chi}_1^\pm \tilde{\chi}_2^0$ associated production which is most promising to detect the SUSY partners of weakly interacting SM particles at the Tevatron. Moreover, the trilepton channel where $\tilde{\chi}_1^\pm \to \ell \nu \tilde{\chi}_1^0$ and $\tilde{\chi}_2^0 \to \ell\ell \tilde{\chi}_1^0$ is considered the *golden mode* due to its striking signature with three leptons and missing transverse energy ($\not{E}_T$). The results of the search are interpreted in mSUGRA with chargino and neutralino masses following the relation $m_{\tilde{\chi}_1^\pm} \approx m_{\tilde{\chi}_2^0} \approx 2 m_{\tilde{\chi}_1^0}$. For large values of tan$\beta$, the $\tilde{\tau}$ becomes the lightest slepton ($\tilde{\ell}$) increasing the branching ratios to final states with 3 taus. The production and the branching ratio also depend on the mass of the sleptons. Several scenarios can be considered: the *3ℓ-max* scenario (heavy squarks and light sleptons) where the leptonic branching fraction is maximally enhanced and the *large $m_0$* scenario (heavy sleptons) where the chargino-neutralino decays are dominated by W/Z exchange resulting in a small leptonic branching ratio. Relaxing scalar mass unification, the cross section is maximal in the limit of large squark masses (*heavy-squarks* scenario).

## 2. CDF analyses

The CDF Collaboration is currently pursuing the search for chargino-neutralino associated production in three different channels: final states with two electrons and a third lepton (e+e+ℓ); final states with two muons and a third lepton (μ+μ+ℓ); and final states with two electrons and an isolated track (ee+track). The e+e+ℓ and μ+μ+ℓ analyses require two central leptons (|η|<1.0) with $p_T$>20 and 8 GeV/c respectively and third electron or muon with $p_T$> 5 GeV/c. The ee+track analysis requires two central electrons with $p_T$>10 and 5 GeV/c respectively and an isolated track with $p_T$>4 GeV/c and therefore it is sensitive to lower $p_T$ leptons that can be produced in τ decays.

The selection criteria are optimized by maximizing S/B where S is the expect signal and B is the background due to SM particles. After requiring 3 candidate leptons, the background is dominated by di-boson production, heavy flavor decays and misidentified leptons. PYTHIA[3] and MADGRAPH [4] event generators, with a detailed simulation of the detector, are used to estimate the acceptance of the SM background. CDF tested the SM prediction in control regions that where defined *a priori* and where a negligible signal contribution is expected. The signal region is analyzed after all the SM background expectations are confirmed. In Fig. 1 the $\not{E}_T$ distribution for the e+e+$\ell$ and μ+μ+$\ell$ analyses are shown. In the ``signal'' region CDF requires $\not{E}_T$ >15 GeV/c$^2$ and other kinematic selection criteria which depend on the decay channel.

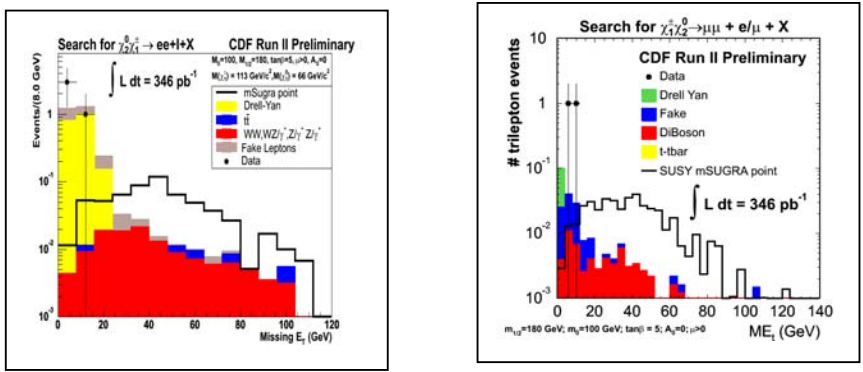

**Figure 1:** $\not{E}_T$ **distribution for the e+e+$\ell$ (left) and μ+μ+$\ell$ (right) analysis.**

The total uncertainty on the signal (background) acceptance is 11% (12%) and 8% (26%) for the trilepton analyses in the muon and electron channel respectively; in the ee+track analysis, the total statistical and systematic uncertainty is 14% (75%) for the signal (background) prediction. In the trilepton analyses, CDF expects 0.09 ±0.03 events in the muon channel and 0.17±0.05 events in the electron channel and observes 0 events in data. In the ee+track analysis, 0.48±0.07 events are expected and 2 events are observed. The estimated number of events from the SUSY signal is 0.37±0.05 and 0.49±0.05 for the muon and electron trilepton analysis, and 0.36±0.27 for the ee+track analysis. The uncertainties quoted are obtained by adding the statistical and systematic errors in quadrature. The expectations are obtained with the next to leading order (NLO) program PROSPINO [5] with *tanβ*= 5, $m_{1/2}$ =180 GeV/c$^2$, $m_0$ = 100 GeV/c2, μ> 0 and $A_0$= 0 followed by the decay of $\tilde{\chi}_1^\pm$ and $\tilde{\chi}_2^0$ with PYTHIA[3]. The luminosity used in these analyses was 346 pb$^{-1}$ for the trilepton and 224 pb$^{-1}$ for the dielectron and track analysis. CDF has not yet set exclusion limits on $\tilde{\chi}_1^\pm \tilde{\chi}_2^0$ associated production. The collaboration is planning to increase the lepton acceptance by including the forward region and by combining the results of these analyses with three other analyses with low momentum μ+μ+$\ell$, high momentum e+μ+$\ell$, and low momentum e+μ+$\ell$ which are ongoing.

### 3. D0 analyses

The D0 collaboration has developed six analyses searching for chargino-neutralino associated production in final states containing three leptons and missing transverse energy: final states with an electron, a tau decaying hadronically, and a third lepton (e+$\tau_{had}$+$\ell$); final states with a muon, a tau decaying hadronically and a third lepton ($\mu$+$\tau_{had}$+$\ell$); final states with two electrons and a third lepton (e+e+$\ell$); final states with an electron a muon and a third lepton (e+$\mu$+$\ell$); final states with two muons and a third lepton ($\mu$+$\mu$+$\ell$); final states with two like sign muons (LS $\mu$+$\mu$). Events are selected with large $\slashed{E}_T$ and two isolated leptons satisfying analysis dependent selection criteria. The analysis containing electrons and muons are described in detail in [6]. D0 has also implemented an analysis focused on the detection of taus decaying hadronically in the final state. The tau leptons are identified by a neural net algorithm which was developed using Z→$\tau\tau$ decays. To maintain the highest possible efficiency the third lepton in all of the six analyses is reconstructed as an isolated track. The isolation conditions for the third track have been designed to be efficient for all lepton flavors, including hadronic tau decay. The dominant SM background sources are dibosons and QCD production.

Table I shows the number of events observed in data and the number of background events expected from Monte Carlo for the six analyses. The number of expected signal events for the m-SUGRA point $m_{\tilde{\chi}_2^0} = 102\,GeV/c^2$, $m_{\tilde{\chi}_1^\pm} = 98\,GeV/c^2$, $m_{\tilde{\ell}_R} = 99\,GeV/c^2$, $tan\beta=3$, $\mu>0$ and $A_0=0$ is also reported in Table 1. The analyses use 320 pb$^{-1}$ of data. The MC has been generated using the CTEQ5L parton distribution functions [8]. The values of $\sigma\times$BF are calculated with CTEQ6L1 at leading order and multiplied with the next to leading order QCD k-factors taken from [5]. The PDF uncertainty of -3.7% to -4.2% is taken into account in deriving limits on the chargino mass.

| Analysis | Data | Background | Expected Signal |
|---|---|---|---|
| e+e+$\ell$ | 0 | 0.21±0.11±0.05 | 3.19±0.14±0.26 |
| e+$\mu$+$\ell$ | 0 | 0.31±0.13±0.03 | 2.17±0.07±0.17 |
| $\mu$+$\mu$+$\ell$ | 2 | 1.75±0.37±0.4 | 1.41±0.04±0.18 |
| LS $\mu$+$\mu$ | 2 | 0.64±0.36±0.13 | 1.03±0.19±0.16 |
| e+$\tau$+$\ell$ | 0 | 0.58±0.11±0.09 | 1.02±0.04±0.08 |
| $\mu$+$\tau$+$\ell$ | 1 | 0.36±0.12±0.06 | 0.65±0.02±0.05 |
| SUM | 4 | 3.85±0.57±0.49 | 9.47±0.50±0.90 |

**Table 1: Number of observed events, SM background and signal events expected in 320 pb$^{-1}$ of data for the D0 trilepton analyses.**

To avoid double-counting, a signal selected by more than one analysis channel is assigned to one channel and removed from the other channels in a way which maximizes the combined sensitivity. No evidence for SUSY has been found. The results are combined to extract limits on the total cross section using the likelihood ratio method (LEP CLs method [7]).

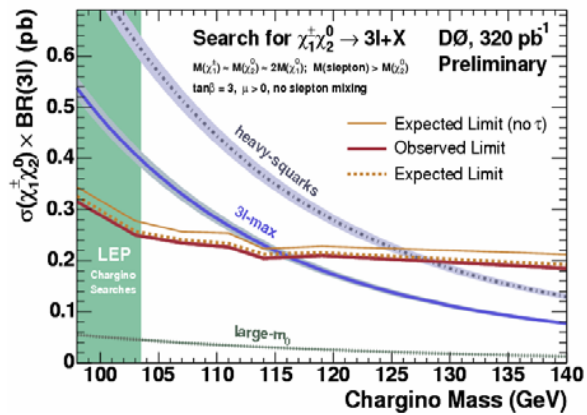

**Figure 2: D0 combined limit on $\sigma\times$BR for $p\bar{p} \to \tilde{\chi}_1^\pm \tilde{\chi}_2^0 \to 3\ell + \slashed{E}_T$ as a function of $m_{\tilde{\chi}_1^\pm}$ for 3 mSUGRA models.**

Assuming the mSUGRA-inspired mass relation $m_{\tilde{\chi}_1^\pm} \approx m_{\tilde{\chi}_2^0} \approx 2m_{\tilde{\chi}_1^0}$ as well as degenerate slepton masses $m(\tilde{\ell})$ (no slepton mixing), the limit on σ×BR(3ℓ) can be derived as a function of $m_{\tilde{\chi}_1^\pm}$ and $m_{\tilde{\ell}}$, with a relatively small dependence on the other SUSY parameters. Therefore, this result can be interpreted in more general SUSY scenarios, as long as the above mass relations are satisfied and R-parity is conserved. The NLO prediction [2] for σ×BR(3ℓ) for the *large m₀, 3ℓ-max* and the *heavy squarks* reference scenarios are shown in Fig. 2. The cross section limit is improved by adding the τ final states and corresponds to a lower limit on the chargino mass of 116 GeV/c² (128 GeV/c²) at 95 % CL in the *2ℓ-max (heavy-squarks)* scenario, which improves the mass limit set at LEP[9]. No limits are set for the *large m₀* scenario.

## 4. Outlook

The sensitivity of the CDF and D0 analyses will improve as the size of the data set collected in Run II will increase. In Fig. 3 we report the combined CDF and D0 limit expectations for run II. The 95% CL limits in the *3-ℓ max (large m₀)* scenario for the chargino mass are 170 GeV (100 GeV), 200 GeV (135 GeV), and 230 GeV (150 GeV) for 2fb⁻¹, 4 fb⁻¹ and 8 fb⁻¹ respectively. This is a very interesting region [10] since there are indications from precision electroweak measurements and astrophysical data that scenarios with light sleptons and squarks might have charginos with masses in the 200 GeV/c² range.

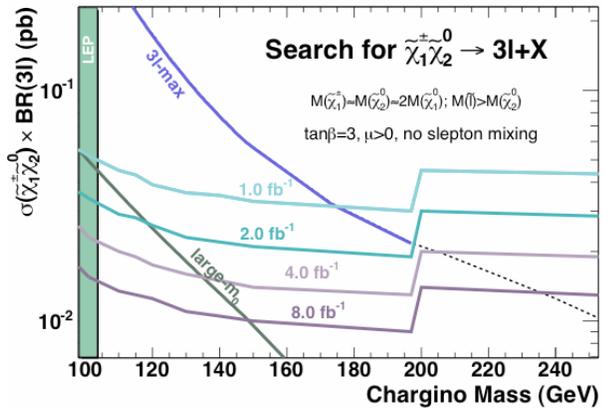

**Figure 3: Expected Tevatron 95% C.L. limit on σ×BR for** $p\bar{p} \to \tilde{\chi}_1^\pm \tilde{\chi}_2^0 \to 3\ell + \not{E}_T$ **as a function of the chargino mass for 2 different mSUGRA models.**